# Observation of Spin Fluctuations in a High-T$_c$ Parent Compound Using Resonant Inelastic X-ray Scattering

B. Freelon[1], P. G. Medaglia[2], A. Tebano[2], G. Balestrino[2], K. Okada[3], A. Kotani[4,5], F. Vernay[6], T. P. Devereaux[6], P. A. Glans[7], T. Learmonth[7], K. E. Smith[7], A. L. D. Kilcoyne[1], B. Rude[1], I. Furtado[1], and J.-H. Guo[1]

[1]*Advanced Light Source, Lawrence Berkeley National Laboratory, One Cyclotron Road, MS 7-100 Berkeley, CA 94720*

[2]*CNR-INFM COHERENTIA and Dipartimento di Ingegneria Meccanica, Università di Roma "Tor Vergata", Via del Politecnico 1, I-00133 Roma, Italy*

[3] *The Graduate School of Natural Science and Technology, Okayama University, 3-1-1, Tsushima-naka, Okayama, Japan 700-8530*

[4]*RIKEN Harima Institute, 1-1-1 Kouto, Sayo, Hyogo 679-5148 Japan*

[5]*Photon Factory, Institute of Material Structure Science, 1-1 Oho, Tsukuba, Ibaraki 305-0801, Japan*

[6]*Department of Physics, University of Waterloo, 200 University Avenue West Waterloo, Ontario, Canada N2L 3G1, & Pacific Institute of Theoretical Physics, University of British Columbia, Vancouver, B.C., Canada V6T 1Z1*

[7]*Department of Physics, Boston University, 590 Commonwealth Avenue, Boston, Massachusetts 02215, USA*

Corresponding Authors: Byron Freelon (bkfreelon@lbl.gov) and Jinghua Guo (jguo@lbl.gov)

Tele: 510-495-2020

FAX: 510 495-2067

Lawrence Berkeley National Laboratory

Advanced Light Source

MS 7-100

Berkeley, CA 94720

We report the first observation of soft-x-ray scattering from spin fluctuations in a high-Tc parent compound.  An antiferromagnetic charge transfer insulator, $CaCuO_2$, was irradiated by Cu $M$-edge soft x-rays.  Ultra-high resolution measurements of scattered intensity revealed magnon-magnon excitations, due to spin exchange scattering, as low-energy loss features.  The process is analogous to optical Raman scattering.  The spectra provide the first measurement of the two-magnon excitation energy and the antiferromagnetic exchange parameter in infinite-layer $CaCuO_2$.  The results reveal resonant inelastic soft x-ray scattering as a novel probe of the spin dynamics in cuprates.

Since its discovery, high-temperature superconductivity (HTSC) has proven to be an immensely complex phenomenon whose microscopic mechanism remains un-deciphered.  This has focused attention on high-Tc parent compounds and the Mott physics associated with these antiferromagnetic (AF) insulators.  The long-range AF order in insulators is quickly destabilized by doping either holes or electrons, giving way to the high- temperature superconducting phase. Upon further doping, the superconducting phase evolves into a region describable in terms of well-defined Landau quasiparticles. The way in which AF correlations and Mott physics declines with doping, allowing high-temperature superconductivity to emerge, is one of the major unsolved problems in condensed matter physics.

AF excitations, or magnons, in parent high-Tc cuprates have been well-observed via inelastic neutron and Raman scattering measurements. While neutron studies have revealed important information like magnon velocities[1], the largest feature observable in Raman scattering measurements is the two-magnon peak due to the scattering of magnons at AF reciprocal lattices vectors $\mathbf{Q}$ and $-\mathbf{Q}$, located at approximately[2] 2700 cm$^{-1}$ or $3J$.  This feature has been observed to soften in energy and intensity with doping, yet remains largely present well into the superconducting phase[3]. However knowledge of the full momentum dependence of the two-magnon spectrum and its evolution with doping would shed important insight into the interplay of Mott physics, AF correlations and high temperature superconductivity.

Resonant Inelastic X-Ray Scattering (RIXS) has rapidly become a powerful technique to detect electronic excitations[4], such as charge-transfer and *d-d* excitations in Mott-physics systems. RIXS provides bulk-sensitive, element-specific, site-selective information, and is applicable in external electric or magnetic fields as it is a photon-in/photon-out process. It has been highly desired to use RIXS for the detection of lower energy excitations like magnons, but limited resolving power of measurements has, to date, made it impossible. In this Letter we report the first x-ray scattering measurements of two-magnon excitations in an AF cuprate. We performed ultra-high resolution experiments on $CaCuO_2$, an AF, high-$T_c$ parent compound exhibiting superconducting behavior in the form of an infinite layer functional block of superlattices[5]. It has been widely considered that the spin dynamics of such systems is strongly related to the high-$T_c$ superconducting pairing mechanism.

By tuning incident photons to absorption resonances, RIXS can provide enhanced signatures[4] of localized and collective low energy excitations (near the Fermi-level) closely linked to antiferromagnetic[6] (AF) order in the cuprates. We note the possibility of magnetic excitations with RIXS along the predicted[7] and demonstrated[8] lines of laser radiation coupling to spin channels. RIXS can be generally considered as a second order Raman effect with scattering intensity given by the Kramers-Heisenberg formula[9]:

$$I\left(\Omega,\omega\right)=\sum_{f}\left|\sum_{i}\frac{\left\langle f\left|\hat{D}_{q'}\right|i\right\rangle\left\langle i\left|\hat{D}_{q}\right|g\right\rangle}{E_{g}+\Omega-E_{i}-i\Gamma_{i}/2}\right|^{2}\delta(E_{g}+\Omega-E_{f}-\omega).$$

We measure spin fluctuations within Cu $3p$ ($M_{2,3}$)-edge scattered intensity where both the ground and final state $|g\rangle$ and $|f\rangle$ respectively, take $3p^6 3d^9$ electron configurations; while the intermediate state $|i\rangle$ is $3p^5 3d^{10}$. $\hat{D}_q$ and $\hat{D}_{q'}$ are dipole operators for the polarizations $q$ and $q'$. $E_g, E_i$ and $E_f$ are the ground, intermediate and final state energies, respectively; and $\Omega$ ($\omega$) are the incident (scattered) photon energies.

The experiments were carried out at Beamline 10.0.1 of the Advanced Light Source in Berkeley. Incident photon energies tuned to the Cu $3p$ absorption edges were calibrated using the absorption resonances of Kr, He and $N_2$ gas with an accuracy of 0.01 eV. RIXS spectra were recorded using a grazing-incidence spectrometer[10]. We used a grating with a groove density of 300 lines/mm and a radius of 3 m. The detector was positioned 90° with respect to the beamline, in the same plane as the beam orbit. The incident polarization vector was parallel to the horizontal and scattering planes. The overall (beamline and spectrometer) experimental resolution was 140 meV.

CaCuO$_2$ thin film samples[11], were grown by pulsed laser molecular beam epitaxy;[12] on NdGaO$_3$ (110) substrates. At room-temperature the orthorhombic films are insulating and possess CuO$_2$ planes perpendicular to the $\hat{c}$-axis. The structural properties of the CaCuO$_2$ were checked by (i) evaluating Bragg-Brentano x-ray diffraction (XRD) patterns and (ii) comparing the x-ray emission spectra of our samples to that of similar materials[13]. XRD indicated excellent crystallographic quality of the thin films. In addition, RIXS is not surface sensitive and the samples were not cleaved. Oxygen-edge x-ray absorption spectra (XAS) of thin film CaCuO$_2$ samples do not contain mobile charge carrier peaks (MCP) that provide a signature[14] of extrinsic hole-doping in the cuprates. The absence of a MCP suggests that there is minimal hole doping due to the growth process or post-growth de-oxygenation. Data was taken at room temperature, in grazing incidence/near-normal take-off mode where the incident photon propagation direction was 20° with respect to the surface plane[15] (a mixture of the $A_{1g}$ and $B_{1g}$ scattering symmetries). This sample position yielded a favorable condition allowing planar CuO scattering, increased penetration depth, and an optimized scattering region observable by the detector[16].

Fig. 1 shows $M$-edge x-ray absorption spectrum of $CaCuO_2$. The absorption profile possesses broad features at the $M_{2,3}$ edges due to the large lifetime broadening of Cu 3p-states and is in reasonable agreement with Cu-$M$ edge absorption spectra from CuO based materials[17]. The vertical hash marks indicate the energies at which RIXS was performed. All of the $M$-edge RIXS spectra contain a truncated elastic - zero energy loss - scattering peak. The data is shown on a vertical scale that enables visualization of spectral weight (features labeled A) to the immediate left of the elastic peak; the intensity of A feature is 1/300 times of the elastic peak. The excitations, located at 390 meV loss, are Raman-like, *i.e.*, the features shift as a linear function of the excitation energies; therefore, the features are stationary on the Raman loss scale as the incident photon energy is varied. Fig. 2 contains magnified regions of scattering intensity profiles resulting from incident energies of 77.5 (a) and 77.6 eV (b) excitation energies in terms of normalized intensity[18]. At loss energies of 1.1-1.3 eV, slight $d_{xy}$ spectral weight is highlighted in the gray field of Fig.1(b). van Veenendaal used the fast-collision approximation to derive $d$-$d$ scattering amplitudes that predicted very weak $d_{xy}$ $M$-edge RIXS spectral weight. There is no A feature intensity for incident polarizations that are not parallel to the $\hat{c}$ -axis. This is in agreement with the claim of a possible novel excitation mode reported[19] in hard x-ray resonant scattering studies on insulating cuprates.

How can one understand the mild shoulder occurring at an energy loss of 0.39 eV? Based on the energy scale, we conclude the A features are neither *CT* nor *d-d* excitations, both of which occur at energy losses greater than ~ 0.5 eV. Such small energy scale features can be understood as a well-known two-magnon excitation due to light scattering from a two-magnetic-sublattice system. The spectral weight of feature A is similar to that suggested[25] as evidence of a spin-pair excitation in the x-ray scattering regime.

To aid in understanding the RIXS spectra, we employed a 12-site single-band Hubbard model illustrated in the inset of Fig. 3(a). We assumed typical parameter values for cuprates: a hopping integral of $t = 0.3$ eV and an on-site Coulomb repulsion of $U = 3.0$ eV, and disregarded the orbital degeneracy for simplicity. The Zhang-Rice singlet band and upper-Hubbard band (UHB) of $CaCuO_2$ were approximated by the lower and upper Hubbard bands of the present single-band Hubbard model, respectively. We also assumed the on-site Coulomb interaction $Q$ between a hole on the Hubbard bands and that on the Cu $3p$ core state (the orbital degeneracy of Cu $3p$ state was also disregarded). We made numerically-exact diagonalization calculations of the Hubbard Hamiltonian to calculate the RIXS spectra under the long x-ray wavelength limit, assuming $\Gamma_i = 0.4$ eV. The spectra were subsequently convoluted with a Lorentz function of FWHM equal to 0.4 eV.

Figure 3(a) shows the X-ray absorption spectrum (XAS) for $Q = 0$ eV and $Q = 3.0$ eV. The $Q = 0$ eV case corresponds to the density of states of the UHB. By increasing $Q$, line shape narrowing occurs. For $Q = 3.0$ eV, the spectrum consists of an almost symmetric main line at 0 eV (the main line position being taken as the origin of the relative photon energy $\Omega$) and a weak and broad satellite structure centered at about 3.0 eV. The energy separation is almost determined by $Q$, indicating the difference in the average valence hole number on the core-hole site; the hole number is almost zero in the final state for the main line, while it is one for the satellite.

The $\Delta\omega = (\omega - \Omega)$ dependence of the RIXS spectrum is shown in Fig. 3(b) for various values of $\Omega$ and for the energy range $0 > \Delta\omega > -1.5$ eV, where the elastic scattering peak at $\Delta\omega = 0$ is omitted. The wide $\Delta\omega$ range plot for $\Omega = 0.3$ eV is shown in the inset of Fig. 3(b). The spectra for $\Delta\omega$ between -4.5 eV and -1.8 eV are assigned to *CT* excitations from the LHB to the UHB. However, the peak at −0.4 eV cannot be interpreted as such a simple inter-band pair excitation. This is due to the two-magnon excitation[25] caused by higher-order perturbation process with respect to $t$. Figure 4 shows an example of the two-magnon excitation mechanism in the RIXS process using a hole (arrow) picture. It is seen from Fig. 3 that the intensity of the two-magnon peak at −0.4 eV reaches a maximum for the incident energy $\Omega \sim 0.3$ eV, about 0.3 eV above the XAS maximum position.

The investigation of the low-energy region for various momenta transfer can be done more accurately in the framework of a spin-only model. The low energy excitation spectra of antiferromagnetic insulators are well described by the simple Heisenberg model. In this spirit, we restrict the scattering operator to spin terms as in numerous Raman studies on AF insulators. The scattering operator $O(q) = \Sigma_{i,\delta} P_{A_{1g}, B_{1g}} S_i S_{i+\delta} \cos(q r_i + q\delta/2)$ is of the Loudon and Fleury type[20] where the operator $P_{A_{1g}, B_{1g}}$ depends on the incident polarization and $\delta$ denotes nearest neighbor vectors[21]. To check that our description is consistent, in the case of $q = 0$, with the RIXS spectrum, we also performed exact-diagonalization on a 16-site cluster with periodic boundary conditions. The scattering intensity is thus given

$$I(q, \Delta\omega) \propto \left| \sum_n \langle \psi_n | \hat{O}(q) | \psi_0 \rangle \right|^2 \delta(\omega - (E_n - E_0)).$$ The results are shown in Fig. 4 where the

total scattering intensity $I$ is the contribution of two symmetries $I \sim I(A_{1g}) + I(B_{1g})$. For $q = 0$, we recover the Raman scattering two-magnon feature around $\sim 2.9J$ which will disperse for finite $q$ to $\sim 4J$ as the Brillouin zone is sampled[21]. A higher excitation exists near $5.5J$ for $q = 0$. These results are consistent with the results of 12-site cluster calculations shown in Fig. 3(b), where the two-magnon peak at $2.9J$ and the higher-order peak at $5.5J$ corresponding, respectively, to 0.4 eV and 0.8 eV features in Fig. 3(b). At higher excitation energies, loss features labeled B exhibit a non-linear Raman shift towards the high energy loss direction (Fig. 1). The B features are near the energy values predicted by the $t$-$J$[22] model and London-Fleury theory (see below) for spin excitations in cuprates due to multiple-magnon scattering excitations involving spins along a chain or within a CuO$_2$ plaquette. Multiple magnon excitations with similar loss energies have been predicted in the pure Heisenberg model with laser Raman scattering. While laser Raman experiments yielding higher order excitations[23] at $\sim 4.4J$ and $5.5J$ influence our current thinking, such work also motivates necessary on-going work to determine the exact nature of the B features in soft x-ray scattering.

de Groot *et al.* succinctly predicted the single spin-flip excitation based on spin-orbit (S.O.) coupling to be symmetry disallowed at the Cu-$M_{2,3}$ edge for cuprates[24] having locally tetragonal symmetry. Our data for the orthorhombic case follows this argument as we observe no features energetically equivalent to a single magnon excitation.

Instead, we observe features at approximately twice the energy of a magnon excitation; similar in energy to two-magnon excitations suggested by Okada and Kotani[25]. Such features were predicted[26] to be discernible from the characteristically small O $K\alpha$ RIXS elastic peaks. While O $K$-edge and Cu $L_{2,3}$-edge RIXS[27] experiments were performed, the two-magnon excitation was not observed due to current spectrometer resolution limitations. In our experiments, the spectral weight of feature A is interpreted as the signature of a two-magnon process occurring via the *exchange scattering mechanism*: the exchange (*c.f.* Fig. 3) of 2 spin states is equivalent to two spins reversals. It is important to contrast the direct observation of the spin-flip excitation through exchange scattering with the indirect methods previously reported. Past observations of spin-flip events, using RIXS, were inferred from changes in resonantly excited *d-d* spectral weight in both $Sr_2CuCl_2O_2$[16] and $NiO$[28]. The observed RIXS spectra allow the superexchange value to be extracted. The energy position of the two-magnon excitation yields a value for the AFM exchange $J$ through the application of the 2D quantum Heisenberg model to $CaCuO_2$. The Heisenberg interaction Hamiltonian is written as $J\sum_{i,j}(\vec{S}_i \bullet \vec{S}_j - \frac{1}{4})$. Such an assumption is reasonable for the insulating cuprates and yields a spin-pair exchange energy of $\omega_{magnon-magnon} \sim 3J$. Thus $J$ is found to equal 0.13 eV; comparable to values (0.11 - 0.14 eV) determined for other cuprates[29].

In summary, using Cu $M$-edge RIXS we have measured two-magnon excitations in the $CaCuO_2$ compound. While the core states of the $CaCuO_2$ are S. O. split, the observed two-magnon excitations are due to spin-exchange scattering. The two-magnon loss energy (390 meV) and the 2D Heisenberg model allows the determination of the CaCuO$_2$ infinite-layer superexchange $J \sim 0.13$ eV. The results show the promise of RIXS as a novel tool for studying *spin dynamics* in high-T$_c$ materials.

The Advanced Light Source is supported by the Director of the Office of Science, Office of Basic Energy Sciences, of the U.S. Department of Energy under Contract No. DE-AC02-05CH11231. T. L. would like to thank the ALS Fellowship Program. T. P. D. acknowledges support from NSERC. The Boston University program is supported, in part, by the Department of Energy under Contract No. DE-FG02-98ERE45680. This work was partially supported by a Grant-in-Aid for Scientific Research from the Ministry of Education, Science, Sports and Culture of Japan.

**Figure List, Freelon *et. al.***

FIG. 1 (a) Cu *M*-edge x-ray absorption spectra of CaCuO$_2$. The open circles (○) represent absorption data while the solid line is a guide to the eye. The vertical hash marks indicate incident photon energies for (b) resonant inelastic x-ray scattering (RIXS) spectra from CaCuO$_2$ plotted on a Raman energy scale. Feature A is a Raman-like two-magnon excitation. Feature B is discussed in the text.

FIG. 2 (a) Cu *M*-edge x-ray emission spectra of CaCuO$_2$ plotted in terms of photon counts versus Raman loss for (a) incident photon energy of 77.5 eV and (b) incident photon energy of 77.6 eV.

FIG. 3 Calculated XAS (a) and RIXS spectra (b). The inset of Fig. (a) shows the unit cell used for the model calculations, for which the periodic boundary condition is assumed along the basis vectors. The inset of (b) shows the RIXS for $\Omega = 0.3$eV in the wide $\Delta\omega$ range where the elastic contribution has been subtracted.

FIG. 4 Schematic representation of two-magnon excitation during resonant core-level excitation and de-excitation. Each solid arrow represents the spin state of a hole. (a) depicts the initial state with AF spin order and (b) shows the intermediate state in which incident radiation creates a core hole at the *i*-th site. This state interacts with the state shown in (c) by exchanging holes. The intermediate state subsequently decays into a final state thereby annihilating the core hole (shown in (d)) and exchanging a hole as the state shown in (e). In the final state (e), we measure the two-magnon excitation.

FIG. 5 Exact diagonalizations of 16-site cluster with periodic boundary conditions. Polarization selecting $A_{1g} + B_{1g}$ symmetry are shown. The relative intensity of the curves is given by the Loudon-Fleury operator.

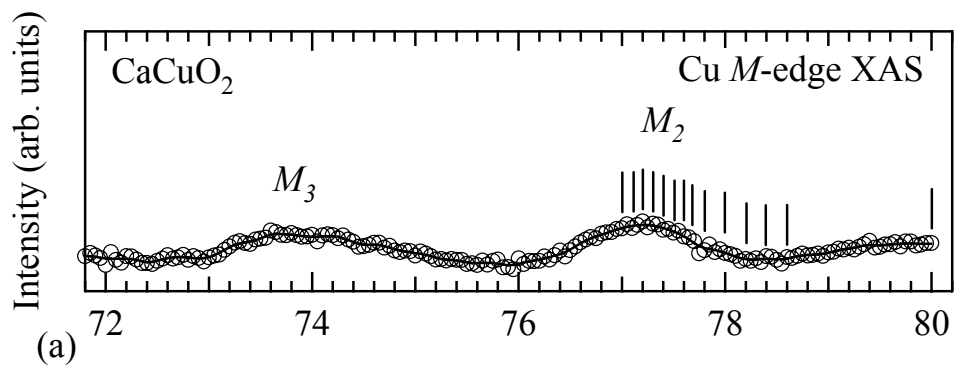

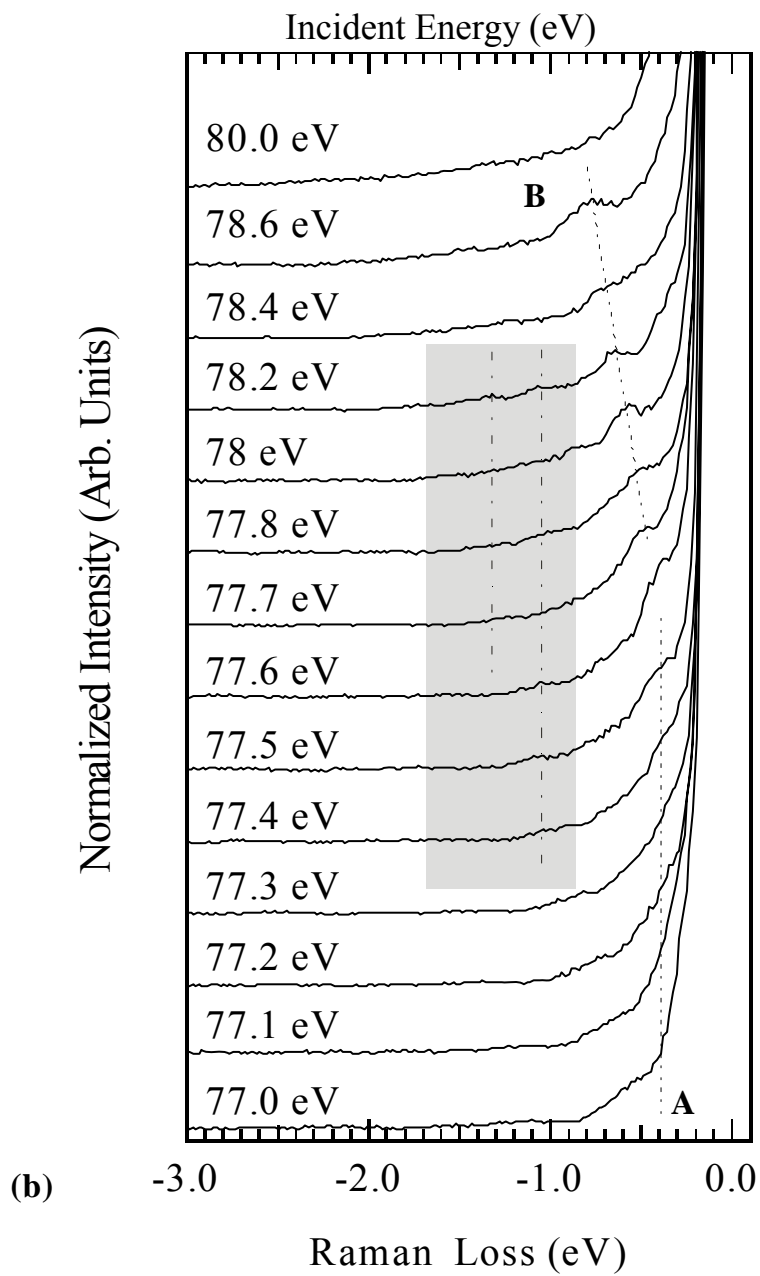

FIG. 1 (a) Cu *M*-edge x-ray absorption spectra of CaCuO$_2$. The open circles (○) represent absorption data while the solid line is a guide to the eye. The vertical hash marks indicate incident photon energies for (b) resonant inelastic x-ray scattering (RIXS) spectra from CaCuO$_2$ plotted in terms of normalized intensity versus a Raman energy scale. Feature A is a Raman-like two-magnon excitation. Feature B is discussed in the text.

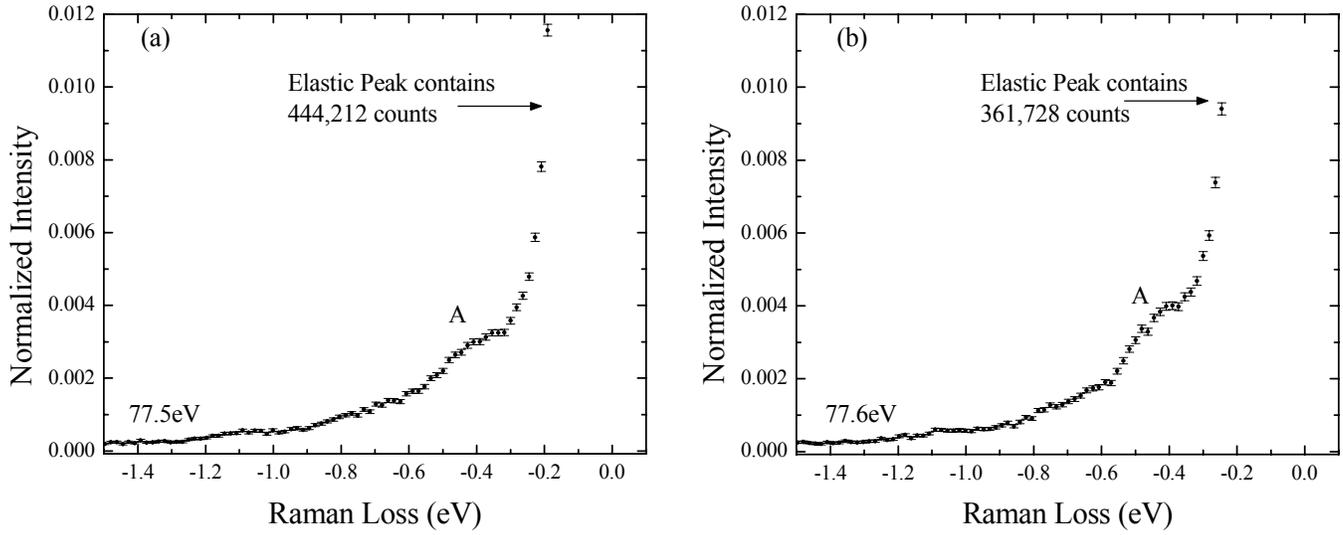

FIG. 2 (a) Cu *M*-edge x-ray emission spectra of CaCuO$_2$ plotted in terms of normalized**Error! Bookmark not defined.** scattered intensity versus Raman loss for incident photon energy of 77.5 eV and (b) incident photon energy of 77.6 eV. All intensity profiles were normalized by setting the elastic peak intensities equal to one.

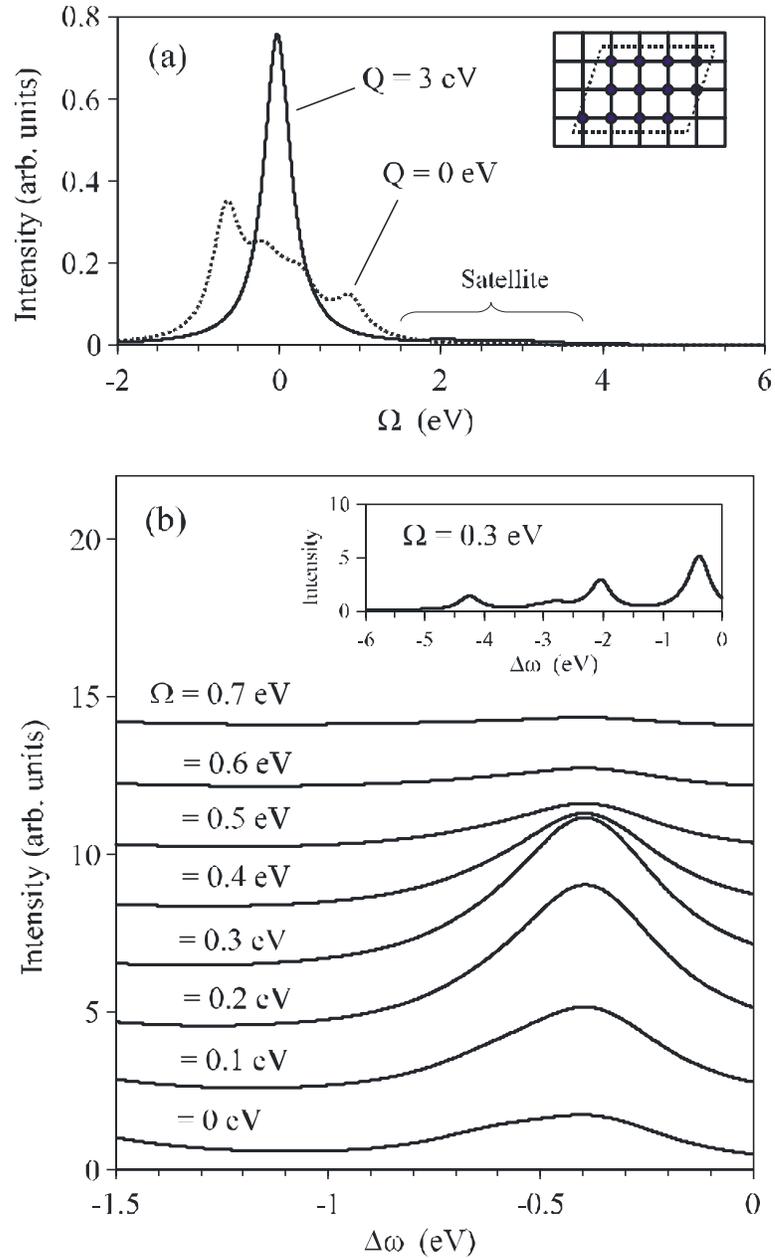

FIG. 3  Calculated XAS (a) and RIXS spectra (b). The inset of Fig. (a) shows the unit cell used for the model calculations, for which the periodic boundary condition is assumed along the basis vectors. The inset of (b) shows the RIXS for $\Omega = 0.3$ eV in the wide $\Delta\omega$ range where the elastic contribution has been subtracted.

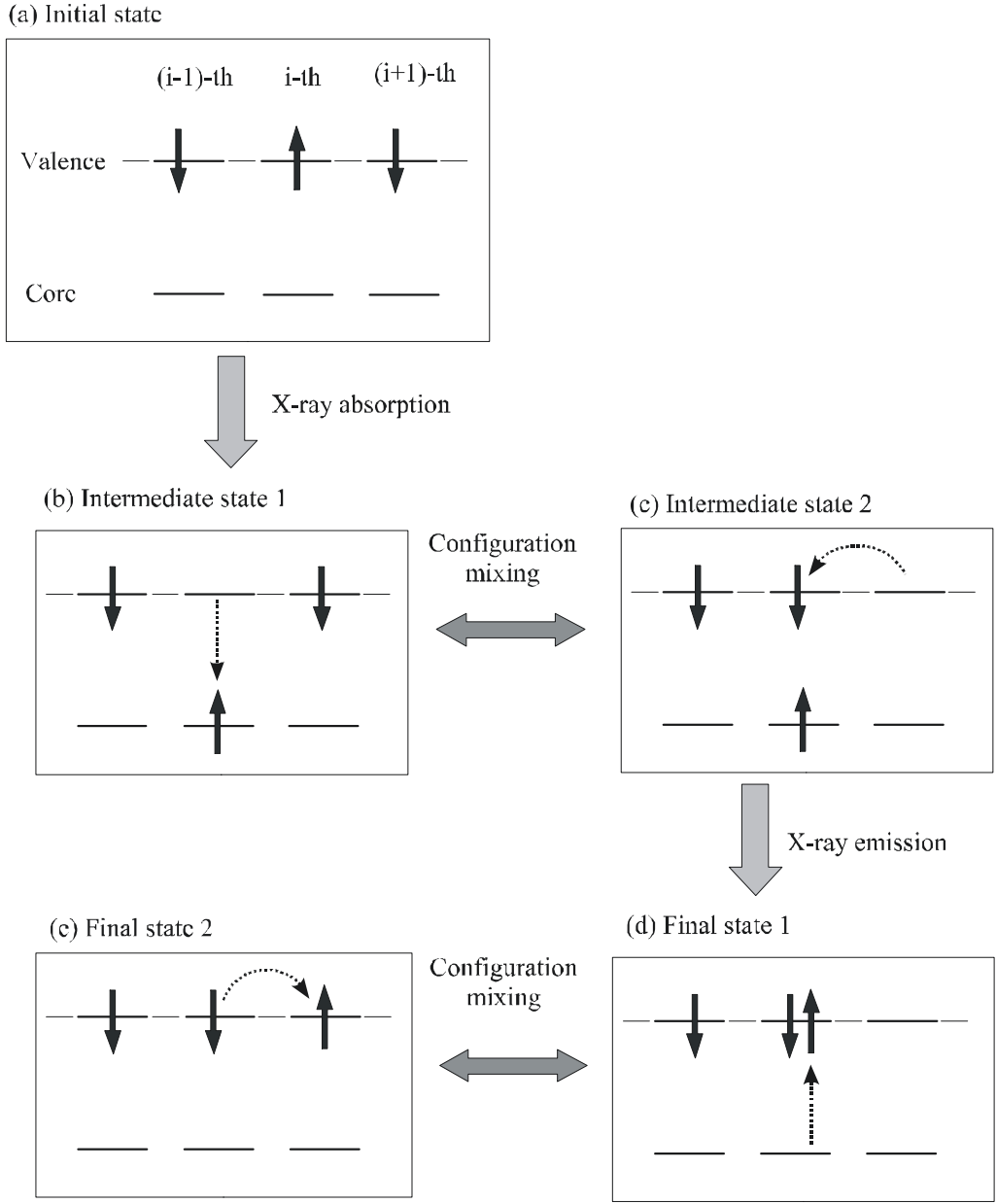

FIG.4 Schematic representation of two-magnon excitation during resonant core-level excitation and de-excitation. Each solid arrow represents the spin state of a hole. (a) depicts the initial state with AF spin order and (b) shows the intermediate state in which incident radiation creates a core hole at the $i$-th site. This state interacts with the state shown in (c) by exchanging holes. The intermediate state subsequently decays into a final state thereby annihilating the core hole (shown in (d)) and exchanging a hole as the state shown in (e). In the final state (e), we measure the two-magnon excitation.

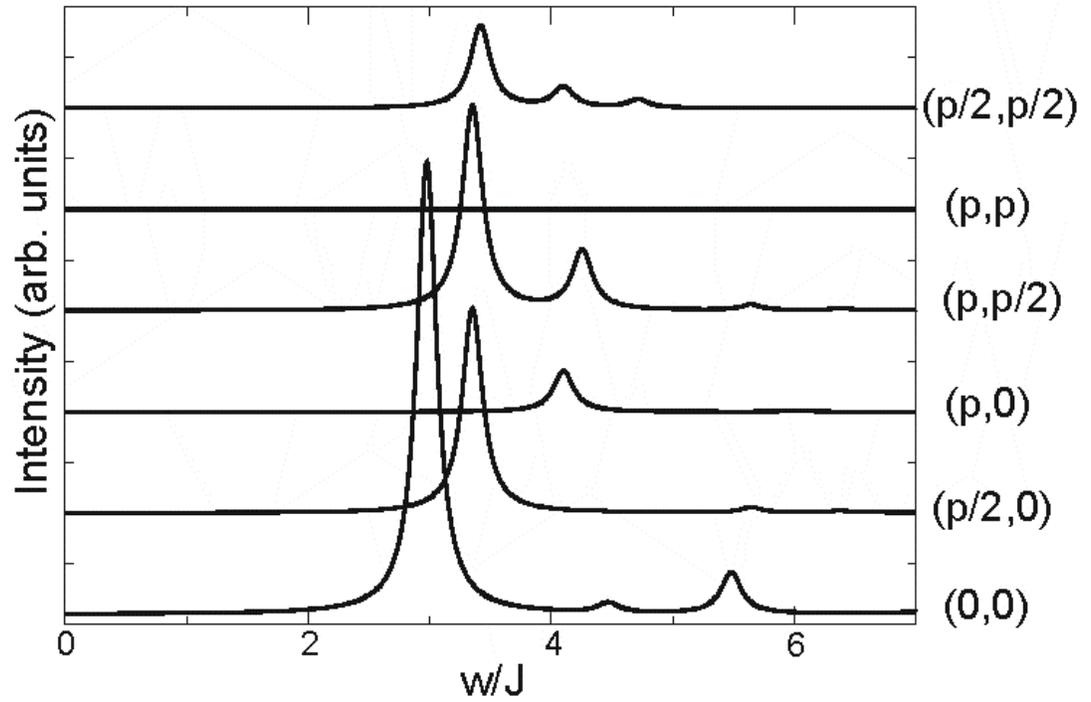

FIG. 5 Exact diagonalizations of 16-site cluster with periodic boundary conditions. Polarization selecting $A_{1g} + B_{1g}$ symmetry are shown. The relative intensity of the curves is given by the Loudon-Fleury operator.